\documentstyle[]{article}

\def\hybrid{\topmargin 0pt      \oddsidemargin 0pt
        \headheight 0pt \headsep 0pt
        \textwidth 6.25in       % A4 paper
        \textheight 9.5in       % A4 paper
        \marginparwidth 0.0in
        \parskip 5pt plus 1pt   \jot = 1.5ex}
\catcode`\@=11
\def\marginnote#1{}

\newcount\hour
\newcount\minute
\newtoks\amorpm
\hour=\time\divide\hour by60
\minute=\time{\multiply\hour by60 \global\advance\minute by-\hour}
\edef\standardtime{{\ifnum\hour<12 \global\amorpm={am}%
        \else\global\amorpm={pm}\advance\hour by-12 \fi
        \ifnum\hour=0 \hour=12 \fi
        \number\hour:\ifnum\minute<10 0\fi\number\minute\the\amorpm}}
\edef\militarytime{\number\hour:\ifnum\minute<10 0\fi\number\minute}

\def\draftlabel#1{{\@bsphack\if@filesw {\let\thepage\relax
   \xdef\@gtempa{\write\@auxout{\string
      \newlabel{#1}{{\@currentlabel}{\thepage}}}}}\@gtempa
   \if@nobreak \ifvmode\nobreak\fi\fi\fi\@esphack}
        \gdef\@eqnlabel{#1}}
\def\@eqnlabel{}
\def\@vacuum{}
\def\draftmarginnote#1{\marginpar{\raggedright\scriptsize\tt#1}}

\def\draftlabel#1{{\@bsphack\if@filesw {\let\thepage\relax
   \xdef\@gtempa{\write\@auxout{\string
      \newlabel{#1}{{\@currentlabel}{\thepage}}}}}\@gtempa
   \if@nobreak \ifvmode\nobreak\fi\fi\fi\@esphack}
        \gdef\@eqnlabel{#1}}
\def\@eqnlabel{}
\def\@vacuum{}
\def\draftmarginnote#1{\marginpar{\raggedright\scriptsize\tt#1}}

\def\draft{\oddsidemargin -.5truein
        \def\@oddfoot{\sl preliminary draft \hfil
        \rm\thepage\hfil\sl\today\quad\militarytime}
        \let\@evenfoot\@oddfoot \overfullrule 3pt
        \let\label=\draftlabel
        \let\marginnote=\draftmarginnote
   \def\@eqnnum{(\theequation)\rlap{\kern\marginparsep\tt\@eqnlabel}%
\global\let\@eqnlabel\@vacuum}  }

\def\numberbysection{\@addtoreset{equation}{section}
        \def\theequation{\thesection.\arabic{equation}}}

\def\underline#1{\relax\ifmmode\@@underline#1\else
        $\@@underline{\hbox{#1}}$\relax\fi}

\def\titlepage{\@restonecolfalse\if@twocolumn\@restonecoltrue\onecolumn
     \else \newpage \fi \thispagestyle{empty}\c@page\z@
        \def\thefootnote{\fnsymbol{footnote}} }

\def\endtitlepage{\if@restonecol\twocolumn \else  \fi
        \def\thefootnote{\arabic{footnote}}
        \setcounter{footnote}{0}}  %\c@footnote\z@ }
\def\nt{non-standard trigonometric }
\def\trig{trigonometric }
\def\trui{trigonometric Ruijsenaars model }

\def\half{{1\over 2}}
\def\sign{\hbox { sign }}

\def\vfc{vertex-face correspondence }
\def\ivs{intertwining vectors }
\def\Rmt{R(u) ^{ij}_{i'j'}}
\def\hsp{\mbox{$\hspace{.5in}$}}
\def\C{{\bf C}}
\def\i{\sqrt{-1}}

\newcommand{\ov}[3]{\phi({#1})^{#2}_{#3}{}}
\newcommand{\tov}[2]{\phi^{#1}_{#2}{}}
\newcommand{\ovell}[3]{\phi^{(ell)}({#1})^{#2}_{#3}{}}
\newcommand{\Ovell}[3]{\varphi^{(ell)}({#1})^{#2}_{#3}{}}
\relax

\numberbysection
\hybrid

\def\beq{\begin{equation}}
\def\eeq{\end{equation}}
\def\p{\partial}

\newdimen\Squaresize \Squaresize=30pt
\newdimen\Thickness \Thickness=0.5pt

\def\Square#1{\hbox{\vrule width \Thickness
   \vbox to \Squaresize{\hrule height \Thickness\vss
      \hbox to \Squaresize{\hss#1\hss}
   \vss\hrule height\Thickness}
\unskip\vrule width \Thickness}    %Example:
\kern-\Thickness}                  %\young{1&2&3&4\cr 5&6&7\cr 8&9\cr 10\cr}

\def\Vsquare#1{\vbox{\Square{$#1$}}\kern-\Thickness}

\begin{document}

\begin{titlepage}

\title{ On trigonometric intertwining vectors and non-dynamical
$R$-matrix for the Ruijsenaars model}

\author{A.Antonov \thanks{Laboratoire
de Physique Th\'eorique et Hautes Energies,
Universit\'e Pierre et Marie Curie, Tour 16 1$^{er}$
\'etage, 4 place Jussieu,
75252 Paris cedex 05-France
and
Landau
Institute for Theoretical Physics
Kosygina str. 2, 117940 Moscow, Russia,
e-mail: antonov@lpthe.jussieu.fr}
\and K.Hasegawa \thanks{Mathematical Institute,
Tohoku University, Sendai JAPAN, e-mail: kojihas@math.tohoku.ac.jp}
\and A. Zabrodin
\thanks{Joint Institute of Chemical Physics, Kosygina str. 4, 117334,
Moscow, Russia and ITEP, 117259, Moscow, Russia,
e-mail: zabrodin@heron.itep.ru}}

\maketitle

\begin{abstract}

We elaborate the trigonometric version of intertwining vectors
and factorized $L$-operators. The starting point is the
corresponding elliptic
construction with Belavin's $R$-matrix. The naive trigonometric
limit is singular and a careful analysis is needed.
It is shown that the construction
admits several different trigonometric degenerations.
As a by-product,
a quantum Lax operator for the trigonometric Ruijsenaars model
intertwined by a non-dynamical $R$-matrix is obtained. The latter
differs from the standard trigonometric $R$-matrix of $A_{n}$ type.
A connection with the dynamical $R$-matrix approach is discussed.

\end{abstract}

\vfill

LPTHE 97-09

\end{titlepage}

\section{Introduction}

Recently, a new type of quantum
Lax operator was suggested \cite{H2}
for the elliptic Ruijsenaars model \cite{R}.
In contrast to the previous versions, this $L$-operator
obeys the standard "$RLL=LLR$" relation with an $R$-matrix
that does not depend on dynamical variables. Specifically,
it is Belavin's elliptic $n^{2}\times n^{2}$
$R$-matrix \cite{B}, where $n$ equals
the number of particles in the model. The construction
relies upon the technique of intertwining vectors and
factorized $L$-operators \cite{Serg}, \cite{H1}, \cite{QF}.

The trigonometric degeneration of this construction is
not automatic since the elliptic intertwining vectors
diverge as the elliptic nome tends to zero. To get a proper
analog, one should either construct
trigonometric intertwining
vectors independently or apply a gauge transformation
before taking the limit.

This paper
deals with that trigonometric limit. As a result, we
get the quantum Lax operator for the \trui which obeys
the $R$-matrix quadratic algebra with a {\it non-dynamical}
trigonometric $R$-matrix. Remarkably, this $R$-matrix
{\it differs} from the standard one \cite{KS}.
In an implicit form the same result was obtained in \cite{ABB}
by means of the {\it dynamical} $R$-matrix approach.

In fact, the trigonometric case allows for different versions
of the intertwining vectors that, therefore, leads to
different types of quantum $L$-operators and
$R$-matrices for the same trigonometric
Ruijsenaars model. Among them, there is a
version without a spectral parameter, which coincides
with the Cremmer-Gervais $R$-matrix \cite{CG}.

Let us give an example of the non-standard
trigonometric $R$-matrix for
$n=2$.
We get it as a twisted degeneration of
Baxter's 4$\times$4 $R$-matrix \cite{Baxter}. It has the form
\beq
R(u)=\left (
\begin{array}{ccccccc}
\sin\, \pi (u+2\eta )&&0&&0&&0\\
&&&&&&\\
0&&\sin\, \pi u&&\sin\, 2\pi \eta&&0\\
&&&&&&\\
0&&\sin\, 2\pi \eta
&&\sin\, \pi u&&0\\
&&&&&&\\
\alpha
\sin\, \pi u \,
\sin\, \pi (u+2\eta )&&0&&0&&\sin\, \pi (u+2\eta )
 \end{array}\right ),
\label{7v}
\eeq
where $\alpha$ is an arbitrary constant. This $R$-matrix
satisfies the Yang-Baxter equation for any
$\alpha$. If $\alpha \neq 0$, we can
set $\alpha =1$ applying a constant gauge
transformation. At $\alpha =0$ we get the ordinary
6-vertex \trig $R$-matrix. By analogy, one may introduce
the {\it 7-vertex} lattice statistical model using
matrix elements of (\ref{7v}) as Boltzmann weights
\cite{ADHR}, \cite{SK}.
For periodic
boundary conditions the partition function
of the 7-vertex
model coincides with that of the 6-vertex one.
The
$R$-matrix (\ref{7v}) appeared for the first time in \cite{Ch}.
Here we present explicit formulas for non-standard
$A_{n-1}$ trigonometric $R$-matrices with spectral parameter for $n\geq 3$.

Particular cases of these $R$-matrices were already mentioned in
the literature \cite{SU}, \cite{ABB}. In \cite{SU}, they were
derived as reductions of
the infinite dimensional $R$-matrix with complete
${\bf Z}$-symmetry (without any relation to the
Ruijsenaars model). In \cite{ABB},
the "modified" commutation relation "$R^*LL=LLR$" for
the trigonometric Ruijsenaars model was pointed out. The
$L$-operator was obtained from the dynamical one
by a gauge transformation depending
on dynamical variables. However, the symmetry
$R=R^*$ of the $R$-matrix leading
to the ordinary relation "$RLL=LLR$" was not noticed there.

Matrix elements of $L$-operators intertwined by the $R$-matrix
(\ref{7v}) obey a quadratic algebra that can be
obtained as a degeneration of the Sklyanin
algebra. In fact, this algebra lies "in between" the
$q$-deformed
universal enveloping
$U_q (sl(2))$ with $q=e^{2i\pi \eta}$ and the
Sklyanin algebra \cite{Skl}.
Representations on this algebra realized
by difference operators were studied in \cite{GZ}. We will see
that the same realization is reproduced by means of trigonometric
intertwining vectors.

Among other applications of the non-standard trigonometric
$R$-matrices we point out their relation to statistical models
of the IRF (interaction round a face) type, in particular,
to "solid-on-solid" (SOS) models. It turns out that
the \nt $R$-matrix is involved in the \vfc
with the \trig SOS model. This can be seen by
constructing
\ivs which are obtained as a special \trig limit
of the elliptic ones. In the elliptic case,
the \vfc
was first established by R.Baxter
in \cite{Baxter} for the eight-vertex model ($n=2$)
and by M.Jimbo, T.Miwa and M.Okado \cite{JMO}
for general $A_{n-1}$-type
models.

Introducing trigonometric intertwining vectors, we then
make the {\it factorized $L$-operator} out of them.
$L$-operators of such a kind first appeared in \cite{IK} for
a particular case.
Later, they were used for the chiral Potts
model and its generalizations \cite{BKMS}. The elliptic version
of factorized $L$-operators was found in
\cite{Serg},
\cite{H1} and \cite{QF}.

The commuting integrals of motion (IM) associated
with this $L$-operator
coincide with Hamiltonian of the trigonometric
Ruijsenaars model. The first
non-trivial Hamiltonian is obtained by taking trace
of the factorized $L$-operator.
Thus, we can refer to this $L$-operator as the
quantum Lax operator for the trigonometric Ruijsenaars model.
This gives a possibility to use the machinery
of quantum inverse scattering method \cite{FT}.
We emphasize this because
other versions of the
\trui $L$-operator satisfies the {\it modified}
Yang-Baxter equation  \cite{GN}, \cite{F}
with {\it dynamical} $R$-matrix \cite{BB},
\cite{ABB}.
These two versions of the \trui
$L$-operators differ by a {\it dynamical} gauge transformation
\cite{ABB}.

This observation might clarify the nature of dynamical $R$-matrices.
We see that there are two $R$-matrix approaches
to the model:
the first one is based on dynamical $R$-matrix \cite{ABB} and
the second one uses the ordinary (non-dynamical) one.
We demonstrate that they lead to the same results.

Note, that another version of quantum elliptic $L$- operator
obtained from cotangent bundles construction was proposed
in \cite{ACF}. In fact, it was connected with factorized
elliptic  $L$- operator \cite{H2} by a twist
transformation with \ivs.

The structure of the paper is as follows. Sect.\,2 contains
a detailed analysis of the $A_1$-case (2$\times$2 $L$-operators).
This corresponds to the 2-particle Ruijsenaars model. Some interesting
algebraic structures related to the Sklyanin algebra are discussed
here. Starting from Sect.\,3 we deal with the general $A_{n-1}$-case
($n$-particle Ruijsenaars model). In Sect.\,3 different
trigonometric $R$-matrices are obtained as
certain degenerations of the elliptic
Belavin $R$-matrix. Different versions of the
trigonometric vertex-face correspondence are discussed in Sect.\,4.
In Sect.\,5 the factorized $L$-operator for
the trigonometric Rujsenaars model is constructed. The
connection with dynamical $R$-matrices is discussed in
Sect.\,6. The Appendices contain some technical remarks.

\section{The twisted trigonometric limit of Baxter's $R$-matrix
and the 7-vertex model}

In this section we discuss a non-standard trigonometric degeneration
of elliptic $R$-matrices for the simplest example of Baxter's
$R$-matrix corresponding to the 8-vertex model.

The universal elliptic $L$-operator with 2-dimensional auxiliary
space has the form
\beq
L(u)=\sum_{a=0}^3 W_{a}(u)S_{a}\otimes \sigma_{a}.
\label{t1}
\eeq
Here
$W_{a}(u)=W_{a}(u|\eta,\tau),\ a=0,\ldots,3$
are functions of the variable $u$
(called the spectral parameter) with parameters $\eta$ and $\tau$:
\beq
W_a(u)={\theta_{\iota (a)}(u)\over
\theta_{\iota (a)}(\eta)}, \;\;\;\;\;\;\;\;
\iota (a) = a+(-1)^a
\label{t2}
\eeq
($\theta_a(x)=\theta _{a}(x|\tau)$ are standard Jacoby theta-functions
with characteristics and the modular parameter $\tau$);
$\sigma_{\alpha}$
are Pauli matrices
($\sigma_0$
is the unit matrix).
The operators
$S_0, S_{\alpha},\ \alpha=1,2,3$, obey the Sklyanin algebra
\cite{Skl}:
\begin{eqnarray}
&&[S_0,S_{\alpha}]_-=iJ_{\beta
\gamma} [S_{\beta},S_{\gamma}]_+,
\nonumber \\
&&[S_{\alpha},S_{\beta}]_-=i[S_0,S_{\gamma}]_+
\label{t4}
\end{eqnarray}
($[A,B]_{\pm}=AB\pm BA$, a triple of Greek indices $\alpha, \beta,
\gamma$ in (\ref{t4}) stands for any {\it cyclic}
permutation of $(1,2,3)$).
Structure constants of this algebra $J_{\alpha \beta}$ have the form
$$
J_{\alpha \beta}={J_{\beta}-J_{\alpha}\over J_{\gamma}} ,
$$
where
$$J_{\alpha}=\frac{\theta _{\iota (\alpha )}(2\eta )
\theta _{\iota (\alpha )}(0)}
{\theta ^{2}_{\iota (\alpha )}(\eta )}$$
Relations (\ref{t4}) were introduced
by E.Sklyanin
as the minimal set of conditions under which
the $L$-operators (\ref{t1})
satisfy the equation
\beq
R_{12}^{(ell)}(u-v)
L_{1}(u) L_{2}(v) = L_{2}(v) L_{1}(u) R_{12}^{(ell)}(u-v),
\label{t5}
\eeq
crucial for the solvability by the algebraic Bethe ansatz method.
Baxter's $R$-matrix $R^{(ell)}(u)$,
\beq
R^{(ell)}(u)=\sum_{a=0}^3 W_{a}(u+\eta)\sigma_{a}\otimes
\sigma_{a},
\label{t6}
\eeq
is the elliptic solution of the quantum Yang-Baxter equation
\beq
R_{12}(u-v) R_{13}(u) R_{23}(v) =R_{23}(v) R_{13} (u) R_{12}(u-v).
\label{t7}
\eeq
In (\ref{t5}), (\ref{t7}) we use the following standard notation:
$L_1 \equiv L\otimes I,\,
L_2 \equiv I\otimes L,$ ($I$ is the identity operator)
The $R$-matrix $R_{12}$ acts
in the tensor product
${\bf C}^2\otimes {\bf C}^2\otimes {\bf C}^2$
as $R(u)$ on the first and the second spaces and as the identity
operator on the third one (similarly for $R_{13}$, $R_{23}$).

Explicitly, the $R$-matrix (\ref{t6}) reads
\beq
R^{(ell)}(u)=\left (
\begin{array}{ccccccc}
a^{(ell)}&&0&&0&&d^{(ell)}\\
&&&&&&\\
0&&b^{(ell)}&&c^{(ell)}&&0\\
&&&&&&\\
0&&c^{(ell)}&&b^{(ell)}&&0\\
&&&&&&\\
d^{(ell)}&&0&&0&&a^{(ell)} \end{array}\right )
\label{t8}
\eeq
(up to a common constant factor),
\begin{eqnarray}
&&a^{(ell)}=\theta _{0}(2\eta |2\tau )
\theta _{0}(u|2\tau )
\theta _{1}(u+2\eta |2\tau ),\nonumber \\
&&b^{(ell)}=\theta _{0}(2\eta |2\tau )
\theta _{1}(u|2\tau )
\theta _{0}(u+2\eta |2\tau ),\nonumber \\
&&c^{(ell)}=\theta _{1}(2\eta |2\tau )
\theta _{0}(u|2\tau )
\theta _{0}(u+2\eta |2\tau ),\nonumber \\
&&d^{(ell)}=\theta _{1}(2\eta |2\tau )
\theta _{1}(u|2\tau )
\theta _{1}(u+2\eta |2\tau ).
\label{t9}
\end{eqnarray}

We are going to study the limit when the elliptic parameter
$h=e^{\pi i \tau }$ tends to $0$. The matrix elements have the
following behavior as $h\rightarrow 0$:
\begin{eqnarray}
&&a^{(ell)}=2h^{1/2}\sin\, \pi (u+2\eta )+O(h^{5/2}),
\nonumber \\
&&b^{(ell)}=2h^{1/2}\sin\, \pi u +O(h^{5/2}),
\nonumber \\
&&c^{(ell)}=2h^{1/2}\sin\, 2\pi \eta +O(h^{5/2}),
\nonumber \\
&&d^{(ell)}=8h^{3/2}\sin\, 2\pi \eta \,
\sin\, \pi u  \,
\sin\, \pi (u+2\eta )+O(h^{11/2}).
\label{t10}
\end{eqnarray}
The limit $h\rightarrow 0$ yields the standard trigonometric
$R$-matrix.

One may apply an $h$-dependent gauge transformation before
taking the limit. Let us choose the matrix of the gauge
transformation in the form
\beq
G=\left ( \begin{array}{ccc}
h^{1/4}\gamma ^{-1/2}&&0\\
&&\\
0&& h^{-1/4}\gamma ^{1/2}\end{array}\right )
\label{t11}
\eeq
with a parameter $\gamma$. This transformation becomes singular
at the limiting point $h=0$, so the result of the limit
appears to be different (though trigonometric)
from the standard trigonometric
$R$-matrix. We set $G_1 = G\otimes I$,
$G_2 = I\otimes G$, as usual, so
$ G_1 G_2 = \mbox{diag}\, \big (
h^{1/2}\gamma ^{-1},\,1, \, 1 ,\,
h^{-1/2}\gamma \big )$. The gauge-transformed $R$-matrix is
$R_h (u)= G_1 G_2 R^{(ell)}(u)(G_1 G_2 )^{-1}$. We define
\beq
R(u)=\frac{1}{2}\lim _{h\rightarrow 0}h^{-1/2}R_h (u).
\label{t12}
\eeq
The result is (up to an irrelevant common factor)
\beq
R(u)=\left (
\begin{array}{ccccccc}
a&&0&&0&&0\\
&&&&&&\\
0&&b&&c&&0\\
&&&&&&\\
0&&c&&b&&0\\
&&&&&&\\
d&&0&&0&&a \end{array}\right ),
\label{t13}
\eeq
where
\begin{eqnarray}
&&a=\sin\, \pi (u+2\eta ),
\nonumber \\
&&b=\sin\, \pi u ,
\nonumber \\
&&c=\sin\, 2\pi \eta ,
\nonumber \\
&&d=4\gamma ^{2}\sin\, 2\pi \eta \,
\sin\, \pi u \,
\sin\, \pi (u+2\eta ).
\label{t14}
\end{eqnarray}

The corresponding gauge transformation of the $L$-operator is
$L(u)\rightarrow GL(u)G^{-1}$. In order to find its explicit
limiting form, it is necessary to fix the behavior of the
operators $S_a$ as $h\rightarrow 0$. Using results of the
paper \cite{GZ}, we define generators $A,\, B, \, C, \, D$
of the degenerate Sklyanin algebra by extracting singular
terms in the expansion of $S_a$ near $h=0$:
\begin{eqnarray}
&&
A=-\, \frac{h^{-1/2}}{2\sin \, 2\pi \eta }
(\cos \,\pi \eta \, S_0 +i\sin \, \pi \eta \, S_3 ),
\nonumber\\
&&
D=-\, \frac{h^{-1/2}}{2\sin \, 2\pi \eta }
(\cos \,\pi \eta \, S_0 -i\sin \, \pi \eta \, S_3 ),
\nonumber\\
&&
C=-\, \frac{h^{1/2}}{2\sin \, 2\pi \eta }
(S_1 -iS_2 ),
\nonumber\\
&&
B=-\, \frac{h^{-3/2}}{8\sin \, 2\pi \eta }
(S_1 +iS_2 ).
\label{t15}
\end{eqnarray}
>From now on we put
$\gamma =1$ without loss of generality.
With this definition we obtain at $h=0$ the following
trigonometric $L$-operator:
\beq
L(u)=\left ( \begin{array}{ccc}
{\cal A}(u)&&{\cal B}(u)\\
&&\\
{\cal C}(u)&&{\cal D}(u)\end{array}\right ),
\label{t16}
\eeq
where
\begin{eqnarray}
&&{\cal A}(u)=e^{i\pi u}A-e^{-i\pi u}D, \nonumber \\
&&{\cal B}(u)=i\sin \, 2\pi \eta \,C, \nonumber \\
&&{\cal C}(u)=2i
\sin \, 2\pi \eta \, \left [2B-(\cos\, 2\pi u \,-\,
\cos \, 2\pi \eta )C\right ], \nonumber \\
&&{\cal D}(u)=e^{i\pi u}D-e^{-i\pi u}A.
\label{t17}
\end{eqnarray}
The generators $A,B,C,D$ satisfy the quadratic
algebra \cite{GZ}:
\begin{eqnarray}
&& DC=e^{2\pi i\eta }CD,\;\;\;\;\;\;CA=e^{2\pi i\eta }AC,
\nonumber \\
&& AD-DA=-2i\,\sin ^{3} 2\pi \eta \, C^2, \nonumber \\
&&BC-CB=\frac{A^2 -D^2 }{2i\sin \, 2\pi \eta }, \nonumber \\
&& AB-e^{2\pi i\eta }BA=e^{2\pi i\eta }DB-BD=
\frac{i}{2}\sin \, 4\pi \eta \,(CA-DC).
\label{algebra}
\end{eqnarray}
The Casimir elements are
\begin{eqnarray}
&&\Omega _{0}=e^{2\pi i \eta }AD -\sin ^{2}2\pi \eta \, C^2,
\nonumber \\
&&\Omega _{1}=\frac{e^{-2\pi i\eta }A^2 +e^{2\pi i \eta }D^2 }
{4\sin ^{2}2\pi \eta }
-BC -\cos \, 2\pi \eta \, C^2 .
\label{casimirs}
\end{eqnarray}
Similar degenerations of elliptic quadratic algebras were considered
in \cite{O}.

An important class of representations of this algebra is given by
the following explicit construction. Let $\Phi (u)$ denote the
$c$-number matrix
\beq
\Phi (u)=\left ( \begin{array}{ccc}
1 &&1 \\
&&\\
2\cos \, \pi (u-2\lambda _{1}) &&
2\cos \, \pi (u-2\lambda _{2})
\end{array} \right ),
\label{t18a}
\eeq
where $\lambda _{1}$, $\lambda _{2}$ are complex parameters.
The factorized $L$-operator can be written as
\beq
L^{(F)}(u)= : \Phi (u+2\ell \eta )
e^{\eta \sigma _{3}(\p _{\lambda _{1}} -\p _{\lambda _{2}})}
\Phi ^{-1}(u-2\ell \eta ):\, ,
\label{t19}
\eeq
where $\ell$ is a constant, $\sigma _{3}= \mbox{diag } (1,-1)$
and the normal ordering :: means that the operators
$\p _{\lambda _{i}}$ should be moved to the right after
performing the matrix product. Explicitly, acting to
functions of $\lambda _{1}$, $\lambda _{2}$, we have:
\begin{eqnarray*}
&&L_{ij}^{(F)}(u)f(\lambda _{1}, \lambda _{2})
\nonumber \\
&=&\Phi _{i1}(u+2\ell \eta )
\left [ \Phi ^{-1}(u-2\ell \eta ) \right ] _{1j}
f(\lambda _{1}+\eta , \lambda _{2}-\eta )
\\
&+&\Phi _{i2}(u+2\ell \eta )
\left [ \Phi ^{-1}(u-2\ell \eta ) \right ] _{2j}
f(\lambda _{1}-\eta , \lambda _{2}+\eta ).
\end{eqnarray*}
It can be shown that for any $\ell$ the $L$-operator
(\ref{t19}) satisfies the intertwining relation (\ref{t5}) with
the $R$-matrix (\ref{t13}).

Therefore, we get a family of representations of the algebra
(\ref{algebra}) realized in the space of functions
$f(\lambda _{1},  \lambda _{2})$. Comparing (\ref{t17}) and
(\ref{t19}), we identify
\beq
L^{(F)} (u+\lambda _{1}+\lambda _{2})=L(u),
\label{t20}
\eeq
where $L(u)$ is as in (\ref{t16}), (\ref{t17}) up to an
irrelevant common factor. This identification gives the following
realization of the algebra (\ref{algebra}) by difference operators:
\begin{eqnarray}
&&A=\frac{e^{-2\pi i \eta \ell }} {\sin \, \pi \lambda _{21}}
\left ( e^{-\pi i \lambda _{21}} T
- e^{\pi i \lambda _{21}} T^{-1} \right ),
\nonumber \\
&&B+\frac{1}{2} \cos \, 2\pi \eta \, C
=\frac{1}{2i \sin \, 2\pi \eta \,
\sin \, \pi \lambda _{21}}
\left ( \cos \, 2\pi ( \lambda _{21} +2 \ell \eta ) T
-\cos \, 2\pi ( \lambda _{21} -2 \ell \eta ) T^{-1}
\right ),
\nonumber \\
&&C
=-\, \frac{1}{i \sin \, 2\pi \eta \,
\sin \, \pi \lambda _{21}}
\left ( T
- T^{-1}
\right ),
\nonumber \\
&&D=\frac{e^{2\pi i \eta \ell }} {\sin \, \pi \lambda _{21}}
\left ( e^{-\pi i \lambda _{21}} T^{-1}
- e^{\pi i \lambda _{21}} T \right ).
\label{t21}
\end{eqnarray}
Here $T^{\pm 1}f(\lambda _{1} , \lambda _{2})=
f(\lambda _{1}\pm \eta , \lambda _{2}\mp \eta )$,
$\lambda _{21}\equiv \lambda _{2} -\lambda _{1}$.

Note that $T$ commutes with functions of $\lambda _{1}+\lambda _{2}$
that allows one to consider $\lambda _{1}+\lambda _{2}$ as a
constant including it in the spectral parameter
After evident redefinitions, the realization (\ref{t21}) coincides
with the one given in \cite{GZ}. The parameter $\ell$ is called
spin of the representation. If $\ell$ is a positive integer or
half-integer, there is a $(2\ell +1)$-dimensional invariant
subspace spanned by symmetric Laurent polynomials of
degree $2\ell$.

Taking trace of the $L$-operator ({\ref{t20}), we get, up to
a common $u$-dependent factor, the operator
\beq
\hat H= \frac{ \sin \, \pi \, (\lambda _{21} +\ell \eta )}
{ \sin \, \pi \, \lambda _{21}}
e^{\eta (\p _{\lambda _{1}} -\p _{\lambda _{2}})}
+\frac{ \sin \, \pi \, (\lambda _{21} -\ell \eta )}
{ \sin \, \pi \, \lambda _{21}}
e^{-\eta (\p _{\lambda _{1}} -\p _{\lambda _{2}})},
\label{t22}
\eeq
which is "gauge equivalent" to the Hamiltonian
of the trigonometric 2-particle Ruijsenaars model.

Let us consider two important limits of the factorized
$L$-operator (\ref{t19}). One of them is to replace
$\lambda _{k} \rightarrow \lambda _{k} +i \Lambda $ and after that
let $\Lambda \rightarrow +\infty$ simultaneously with the
gauge transformation $L(u) \rightarrow g(u)L(u) g^{-1}(u)$,
where $g(u)$ is the diagonal matrix
$$
g(u)= \mbox { diag } \, \left (
\exp [ \frac{i \pi }{2}(u-2i \Lambda )],\,\,
\exp [ -\frac{i \pi }{2}(u-2i \Lambda )]
\right ).
$$
In this way we get the {\it $u$-independent $L$-operator}
\beq
L=\frac{1}{\sin \, \pi \lambda _{21} }
\left (
\begin{array}{ccc}
e^{-2i \pi (\lambda _{2}+\ell \eta )}T
-e^{-2i \pi (\lambda _{1}+\ell \eta )}T^{-1}
&& T^{-1}-T
\\
&&\\
e^{-2i \pi (\lambda _{1}+\lambda _{2} )}(T-T^{-1}) &&
e^{-2i \pi (\lambda _{2}-\ell \eta )}T^{-1}
-e^{-2i \pi (\lambda _{1}-\ell \eta )}T
\end{array}
\right )
\label{t23}
\eeq
This $L$-operator is intertwined by the trigonometric $R$-matrix
\beq
R'(u)=\left (
\begin{array}{ccccccc}
\sin\, \pi (u+2\eta )&&0&&0&&0\\
&&&&&&\\
0&&\sin\, \pi u&&e^{-\pi i u}\sin\, 2\pi \eta&&0\\
&&&&&&\\
0&&e^{\pi i u}\sin\, 2\pi \eta
&&\sin\, \pi u&&0\\
&&&&&&\\
0&&0&&0&&\sin\, \pi (u+2\eta )
\end{array}\right ),
\label{t24}
\eeq
i.e., it holds
\beq
R' _{12}(u-v)L_1 L_2 = L_2 L_1
R' _{12}(u-v).
\label{t25}
\eeq
Note that the $R$-matrix {\it depends} on the spectral parameter
while the $L$-operators do not. Tending $u-v \rightarrow i\infty$,
we get the version of these commutation relations
without spectral parameter.

Let us stress that $\mbox{tr} L$ again yields the trigonometric
Ruijsenaars Hamiltonian (\ref{t22}). At the same time
commutation relations (\ref{t25}) define the algebra of functions
$\mbox{Fun}_{q}(SL(2))$
on the quantum group $SL(2)$ with
$q=\exp (2\pi i \eta )$. The $L$-operator (\ref{t23}) provides
a representation of this algebra. Therefore, the 2-particle
trigonometric Ruijsenaars model appears to be connected with
representation of the algebra
$\mbox{Fun}_{q}(SL(2))$.

Another limit yields the $L$-operator constructed from
representations of the {\it dual algebra} to  the quantum
algebra of functions
$\mbox{Fun}_{q}(SL(2))$, the $q$-deformation of the
universal enveloping of the $sl(2)$ algebra,
$U_{q} (sl(2))$.
The limit consists in replacing
$\lambda _{1} \rightarrow \lambda _{1} -i \Lambda $,
$\lambda _{2} \rightarrow \lambda _{2} +i \Lambda $
simultaneously with a proper gauge transformation such that the
limit $\Lambda \rightarrow \infty$ is finite.
In this way we get the $L$-operator
\beq
L(u) \rightarrow
\left (
\begin{array}{ccc}
e^{i \pi (u- 2\ell \eta )}T
-e^{-i \pi (u-2\ell \eta )}T^{-1}
&& e^{i \pi \lambda _{21}} (T^{-1}-T)
\\
&&\\
e^{-i \pi \lambda _{21}}(e^{-4i \pi \ell \eta }T-
e^{4i \pi \ell \eta }T^{-1}) &&
e^{i \pi (u+2\ell \eta )}T^{-1}
-e^{-i \pi (u+2\ell \eta )}T
\end{array}
\right ).
\label{t26}
\eeq
This $L$-operator is intertwined by the
standard trigonometric $R$-matrix. Its trace is an operator
with constant coefficients.

The fundamental representation of the
algebra (\ref{algebra}) acts in the 2-dimensional
invariant subspace for $\ell =\frac{1}{2}$.
Explicitly, it is provided by
\beq
L(u)=R(u-\eta ).
\label{t18}
\eeq

Let us introduce a lattice statistical model using the $R$-matrix
(\ref{t13}) as the matrix of Boltzmann weights
corresponding to
different configurations
of arrows around a vertex.
We call it {\it 7-vertex model} because there are
7 non-zero weights. Note that the matrix
elements $a,b,c,d$ are independent Boltzmann weights.
They always can be parameterized as in (\ref{t14}).
The 6-vertex model is
reproduced at $\gamma =0$.

The monodromy matrix is constructed in the standard way:
\beq
{\cal T}(u)=R_{0N}(u)\,\ldots \, R_{02}(u)R_{01}(u)
\label{7.1}
\eeq
(the auxiliary "horizontal" space is labeled by $0$).
The transfer matrix is
$
T(u)=\mbox{tr}_{0}{\cal T}(u)
$,
where trace is taken in the auxiliary space. It follows from the
Yang-Baxter equation that $[T(u),\, T(v)]=0$.
The partition function
for the toroidal $M\times N$ lattice is
$
{\cal Z}=\mbox {Tr}\, T^M (u)
$
(here the trace is taken in the tensor product of vertical
spaces).

It is well known that in
any configuration of arrows on the
toroidal lattice the $d$-vertices always come in pairs with
the $d'$-vertices (in which all arrows are reversed), so the
partition function depends on $(dd')^{1/2}$ only. In our case
this means that the partition function does not depend on
$\gamma$, i.e. it is the same as that of the standard 6-vertex
model. However, the eigenvectors of the transfer matrix are
different. In particular, for non-zero $\gamma$ the transfer
matrix is not completely diagonalizable,
i.e., it contains Jordan cells.
Since the number
of vertices looking up and down along the row is not conserved
from row to row, the
usual Bethe ansatz technique for finding eigenvectors
is not applicable. One should apply Baxter's method used
in the 8-vertex model. Whence the 7-vertex case
might serve as a
toy model of Baxter's elliptic construction.

\section{Non-standard trigonometric $R$-matrices for $n\geq 3$}

In this section we obtain the non-standard trigonometric
degenerations
of the elliptic Belavin $R$-matrix \cite{B}. To write it
down explicitly,
we need theta functions with rational characteristics
\beq
\theta^{(j)}(u)
= \sum_{m \in {\bf Z}}
\exp \left [ \pi \i n \tau \big (m+\half-\frac{j}{n}\big )^2 +
2\pi \i \big (m+\half-\frac{j}{n}\big )(u+\half)\right ].
\label{thetaj}
\eeq
(Here and below we write $\i$ instead of the imaginary unit to avoid
coincidence with the index label.)
Choosing the standard bases
$(E_{ii'})_{kl} =\delta _{ik}\delta _{i'l}$
in the space of $n^{2}\times n^{2}$ matrices,
we write the Belavin $R$-matrix in the form
$$
R^{(ell)}(u)
=
\sum_{i',j'=1}^n  R^{(ell)}(u) ^{i'j'}_{ij}
E_{ii'}\otimes E_{jj'},
$$
where the matrix elements are \cite{RT}:
\begin{equation}
R^{(ell)}(u) ^{ij}_{i'j'} =
\delta_{i+j, i'+j' {\rm mod} n}
\frac
        {\theta^{(i'-j')} (u+2\eta) \theta^{(0)}(u)}
        {\theta^{(i'-i)} ({2\eta})\theta^{(i-j')}(u)}.
\label{brm}
\end{equation}
The normalization here is different from the one used in the
previous section for $n=2$.

As in the case $n=2$,
we consider the limit when the elliptic
nome $h=e^{\pi i \tau}$ tends to 0.
The theta functions
$\theta^{(j)}(u)$ behave as
\begin{eqnarray*}
&&\theta^{(j)}(u)
= \! \exp \left [ 2\pi \i \big (
\frac{1}{2} \mbox{sign}\, j -\frac{j}{n}
\big )\! \big ( u+\frac{1}{2} \big )\right ]
h^{n(\half-\frac{|j|}{n})^2}
+O \! \left (
h^{n(\half+\frac{|j|}{n})^2}\right ),
\;\;\;\; -n+1 \leq j \leq n-1, \;\;\; j\neq 0,
\nonumber \\
&&\theta^{(0)}(u)
= -2 \sin \pi u\, h^{n/4}
+ O \left ( h^{\frac{9}{4}n} \right ).
\end{eqnarray*}
So the asymptotic of the Belavin
$R$-matrix $R^{(ell)}(u) ^{ij}_{i'j'}$
can be written as follows:
\beq
R^{(ell)}(u)^{ij}_{i'j'} = \delta_{i+j, i'+j' \, {\rm mod}\, n}
\, O \left ( h^{ \frac{1}{n} [ 2(i-j' )(i' -i ) +n |i' -i|
+n|i-j' | -n|i' - j' |]} \right ), \;\;\;\;\;\;h\rightarrow 0.
\label{belass}
\eeq
Taking the limit $h\rightarrow 0$,
we get the standard trigonometric $A_{n-1}$
$R$-matrix \cite{KS}
\begin{eqnarray}
R_{A_{n-1}}(u) ^{ij}_{i'j'}
&=&
\delta _{i,j} \delta_{i,i'} \delta _{i',j'}
\frac{\sin\pi(u+2\eta)}{ \sin 2\pi \eta }
\nonumber \\
&+&\delta_{i,i'} \delta _{j,j'} \,
\varepsilon (i'\neq j')
\frac{\sin \pi u}{ \sin 2\pi \eta}
\exp \left [ \frac{ 2\pi \i \eta }{n}
\big ( 2 (j-i) - n\sign (j-i)\big ) \right ]
\nonumber \\
&+&\delta _{i,j'} \delta _{i',j} \,
\varepsilon (i'\neq j')
\exp \left [ -\frac{ \pi \i u }{n}
\big ( 2(j-i) - n\sign (j-i) \big ) \right].
\label{srm}
\end{eqnarray}
Here, $\varepsilon (\hbox{ condition })$ is equal to 1 if the
condition is true 0 otherwise.

As in Sect.\,2, we may apply an $h$-dependent gauge transformation
before the limit $h \rightarrow 0$.
The matrix of gauge transformation has the form
\beq
G_{ij}= \delta _{i j}  h^{n(\half-\frac{i}{n})^2},
\hsp i,j= 1, \ldots, n.
\label{gaugmat}
\eeq
The result of the limit differs from the
standard trigonometric $A_{n-1}$ $R$-matrix because the gauge
transformation is singular at the point $h=0$.

We are interested in the non-standard trigonometric limit
\beq
\tilde R(u)=\lim _{h\rightarrow 0}
G_1 G_2 R^{(ell)}(u) G_1^{-1}  G_2 ^{-1},
\label{lim}
\eeq
where
$G_1 = G\otimes I$,
$G_2 = I\otimes G$.

Matrix elements of
the Belavin $R$-matrix \ref{brm}) are non-zero provided
$i+j=i'+j'+ M$ for $M=-n, 0, n$. It is not difficult to see that
the leading terms
of the gauge transformed elliptic $R$-matrix as $h \rightarrow 0$
are
\beq
\left(G_1 G_2 R_B(u)G_1^{-1} G_2 ^{-1}\right) ^{ij}_{i'j'}
=
\delta_{i+j, i'+j'\, {\rm mod} \, n}
\, O\left ( h^{ \frac{M}{n}(2j - M)-M+
|i'-i|+ |i-j'| +|i'-j'|
}\right ).
\label{ass}
\eeq
Extracting terms of zero degree in $h$, we get
the non-standard trigonometric
$R$-matrix (\ref{lim}):
\beq
\tilde \Rmt= R_{A_{n-1}}(u)^{ij}_{i'j'} + S(u)^{ij}_{i'j'},
\label{nrm}
\eeq
where
\begin{eqnarray}
S(u)^{ij}_{i'j'}=
&-&
2 \i \sin \pi u \,
\delta_{i+j,i'+j'}\,\varepsilon (i'<i<j')\,
\exp \left [ \frac{
2\pi \i }{n}\big (
(i-i')u + 2 (j' -i )\eta \big )\right ]
\nonumber \\
&+&2 \i \sin \pi u\,
\delta_{i+j,i'+j'}\,\varepsilon (i'>i>j')\,
\exp \left [ \frac{
2\pi \i }{n}\big (
(i-i')u + 2 (j' -i )\eta \big )\right ]
\nonumber \\
&-&2 \i \sin \pi u\,
\delta_{i+j,i'+j'-n}\,\delta_{i',n}\,\varepsilon (j'<n )\,
\exp \left [ \frac{
2\pi \i }{n}\big (
i u + 2j \eta \big ) \right ]
\nonumber \\
&+&2 \i \sin \pi u\,
\delta_{i+j,i'+j'-n}\,\delta_{j',n}\,\varepsilon (i'<n )\,
\exp \left [ \frac{
-2\pi \i }{n}\big (
j u + 2i \eta \big ) \right ]
\nonumber \\
&+&4 \sin \pi u\,
\sin \pi(u+2\eta)\,
\delta_{i+j,i'+j'-n}\,\delta_{i',n}\,\delta_{j',n}\,
\exp \left [\frac{
2\pi \i }{n}\big ( i- \frac{n}{2} \big )
\big ( u-2\eta \big )\right ].
\label{S}
\end{eqnarray}
The classical limit of this $R$-matrix agrees with the Belavin-Drinfeld
classification of classical $r$-matrices \cite{BD}.

The matrix elements with $i+j=i'+j'-n$
(the last 3 lines in (\ref{S})) can be eliminated by
the gauge transformation with the diagonal matrix
\beq
\label{D}
D_{ij}=\delta _{ij} \exp \big [ \pi (n-2j) \Lambda \big ]
\eeq
and subsequent limit $\Lambda \rightarrow -\infty$, so
only the terms with $i+j =i' +j'$ survive. At the same time this
gauge transformation does not
change the form of the Yang-Baxter equation.

As a result, one obtains the $R$-matrix ( appeared in \cite{SU}, \cite{ABB})
\begin{eqnarray}
R(u) ^{ij}_{i'j'}
&=&
\delta _{i,j} \delta_{i,i'} \delta _{i',j'}
\frac{\sin\pi(u+2\eta)}{ \sin 2\pi \eta }
\nonumber \\
&+&\delta_{i,i'} \delta _{j,j'} \,
\varepsilon (i'\neq j')
\frac{\sin \pi u}{ \sin 2\pi \eta}
\exp \left [ \frac{ 2\pi \i \eta }{n}
\big ( 2 (j-i) - n\sign (j-i)\big ) \right ]
\nonumber \\
&+&\delta _{i,j'} \delta _{i',j} \,
\varepsilon (i'\neq j')
\exp \left [ -\frac{ \pi \i u }{n}
\big ( 2(j-i) - n\sign (j-i) \big ) \right]
\nonumber \\
&-&
2 \i \sin \pi u \,
\delta_{i+j,i'+j'}\,\varepsilon (i'<i<j')\,
\exp \left [ \frac{
2\pi \i }{n}\big (
(i-i')u + 2 (j' -i )\eta \big )\right ]
\nonumber \\
&+&2 \i \sin \pi u\,
\delta_{i+j,i'+j'}\,\varepsilon (i'>i>j')\,
\exp \left [ \frac{
2\pi \i }{n}\big (
(i-i')u + 2 (j' -i )\eta \big )\right ].
\label{nrm1}
\end{eqnarray}
Note that for $n=2$ it coincides with the
ordinary 6-vertex $R$-matrix.

\section {The vertex-face correspondence for $n\geq 3$}

Like in the case $n=2$, matrix elements of the
$R$-matrix (\ref{nrm}) can be considered as
Boltzmann weights of a lattice vertex model.
We establish the correspondence of this vertex model with
the trigonometric SOS model.

The trigonometric SOS model is an IRF-type
model on a two-dimensional square lattice with statistical
variables taking values in $\C^n$.
Fix an orthonormal basis
in $\C^n$:
$\C^n=
\oplus_{i=1,\ldots , n}\C \epsilon_i$,
$<\epsilon_i,\epsilon_j>=\delta_{i,j}$ and let
$\bar{\epsilon}_k= \epsilon_k -
\frac{1}{n}\sum_{i=1}^{n} \epsilon_i$ for
$k=1, \ldots , n$.

The Boltzmann weight corresponding to the configuration
$\lambda, \mu, \mu', \nu \in \C^n$ round a face is denoted by
$$
W \left[\begin{array}{ccc}
                        & \mu&\\
                \lambda &u&\nu\\
                        & \mu'& \\\end{array}\right ].
$$
The face weights for admissible configurations\footnote{Usually
the statistical
variables of SOS models take values in the weight space of the
$sl(n)$ algebra realized as the orthogonal complement to the
vector $\sum _{i=1}^{n} \epsilon _{k}$ in $\C^n$. However,
for our purposes it is more convenient to extend the definition
of Boltzmann weights and intertwining vectors to the whole space
$\C^n$. As it is seen from (\ref{facew1})-(\ref{facew3}),
(\ref{vfc}), the
selection rule for admissible weights and
the vertex-face correspondence remain the same.}
are:
\beq
W
\left[\begin{array}{ccc}
                        & \lambda +2\eta\bar{\epsilon_{r}} & \\
                \lambda &u& \lambda + 4\eta\bar{\epsilon_{r}} \\
                        & \lambda +2\eta\bar{\epsilon_{r}} &
                        \\\end{array}\right
]
= \frac {\sin \pi(u+{2\eta})}{\sin {2\pi \eta} },
\label{facew1}
\eeq
\beq
W
\left[\begin{array}{ccc}
                & \lambda +2\eta\bar{\epsilon_{r}} & \\
        \lambda
        &u&\lambda+2\eta(\bar{\epsilon_{r}}+\bar{\epsilon_{s}})\\
                & \lambda +2\eta\bar{\epsilon_{r}} &
                \\\end{array}\right]
= \frac{\sin \pi (-u+\lambda_{rs})}{\sin \pi \lambda_{rs}},
\;\;\;\; r\neq s,
\label{facew2}
\eeq
\beq
W
\left[\begin{array}{ccc}
                & \lambda +2\eta\bar{\epsilon_{r}} & \\
        \lambda &u&\lambda + 2\eta(\bar{\epsilon_{r}}
        +\bar{\epsilon_{s}})\\
                & \lambda +2\eta\bar{\epsilon_{s}} &
                \\\end{array}\right]
= \frac{\sin \pi u}{\sin {2 \pi \eta}}
        \frac{\sin \pi({2\eta}+\lambda_{rs})}
{\sin \pi \lambda_{rs}},
\;\;\;\; r\neq s,
\label{facew3}
\eeq
where
$
\lambda_{rs}\equiv <\lambda, {\epsilon_r} - {\epsilon_s}>$.
For all other configurations of $\lambda, \mu, \mu' ,\nu$
the face weight is put equal to 0.
The weights (\ref{facew1})-(\ref{facew3})
satisfy the star-triangle relation that is an analog of the
Yang-Baxter equation for models of IRF type.

To get the \vfc we introduce the trigonometric
intertwining vectors
$\tilde \ov     {u}
        {\mu}
        {\lambda}
        _ { }$ with components
$\tilde \ov     {u}
        {\mu}
        {\lambda}
        _ {\, j }$, $j=1, \ldots , n$. These vectors depend
on $\lambda , \mu \in \C^n$ as well as on the spectral parameter
$u$. The vectors are set to be zero unless
$\mu - \lambda =2\eta \bar \epsilon _{k}$. Components of the
non-zero vectors are given by
\begin{eqnarray}
&&
\tilde \ov     {u}
        {\lambda + 2\eta \bar \epsilon_k}
        {\lambda}
        _ { j }
=\exp  \left [ \frac{
\pi \i }{n}(n-2j)(u-n<\lambda,{\epsilon}_k>) \right ],
\;\;\;\;\; j= 1, \ldots , n-1,
\nonumber \\
&&
\tilde\ov     {u}
        {\lambda + 2\eta \bar \epsilon_k}
        {\lambda}
        _ { n }
=2\cos \, \pi (u-n<\lambda,{\epsilon}_k>).
\label{trint}
\end{eqnarray}
Then the trigonometric \vfc holds in the form
\begin{equation}
\sum_{i,j=1}^n
\tilde R(u-v)^{i,j}_{i',j'}\,\,
\tilde\ov     {u}
        {\mu}
        {\lambda}
        _{i}
\,\,
\tilde\ov     {v}
        {\nu}
        {\mu}
        _{j}
=
\sum_{\mu'}
\tilde\ov     {v}
        {\mu'}
        {\lambda}
        _{j'}
\,\,
\tilde\ov     {u}
        {\nu}
        {\mu'}
        _{i'}
\,\,W
\left[\begin{array}{ccc}
                        & \mu & \\
                \lambda & u-v &\nu\\
                        & \mu'& \\\end{array}\right]
\label{vfc}
\end{equation}
(see Appendix A).
Here $\tilde  R(u)^{i,j}_{i',j'}$ is the non-standard
trigonometric $R$-matrix (\ref{nrm}), (\ref{S}).

Eq.\,(\ref{vfc}) is the most general form of the trigonometric
vertex-face correspondence. It admits certain simplifications.
Let us shift the vector
$\lambda$:
$$
\lambda\rightarrow\lambda+ \i \Lambda \sum_{k=1}^{n} \epsilon_k \, ,
$$
where $\Lambda$ is a constant.
Then \ivs (\ref{trint})
change as follows:
$$
\tilde\ov     {u}
        {\lambda + 2\eta \bar\epsilon_k}
        {\lambda}
        _ { j }
\rightarrow
\exp \left [ \pi \,(n- 2j)\Lambda \right ]
\tilde\ov     {u}
        {\lambda + 2\eta \bar\epsilon_k}
        {\lambda}
        _ { j }
\;\;\;\;
\hbox { for } j= 1, \ldots, n-1\, ,
$$
$$
\tilde\ov     {u}
        {\lambda + 2\eta \bar\epsilon_k}
        {\lambda}
        _ { n }
\rightarrow
\exp \big (-\pi n \Lambda \big )
\left \{
\exp \left [
-\pi \i (u-n<\lambda,{\epsilon}_k> \right ]+
e^{2\pi n\Lambda }\,
\exp \left [
\pi \i (u-n<\lambda,{\epsilon}_k >\right ] \right \}.
$$
Let us apply to eq.\,(\ref{vfc}) the gauge transformation
with the diagonal matrix (\ref{D}).
Then the $R$-matrix
$\tilde \Rmt$
transforms as
$$
\tilde\Rmt\rightarrow \tilde\Rmt \, \exp
\big [ 2\pi (i'+j'-i-j) \Lambda \big ].
$$
Finally, take the limit $\Lambda \rightarrow -\infty$.
The $R$-matrix $\tilde \Rmt$ turns into
$\Rmt$ (\ref{nrm1}). The intertwining vectors (\ref{trint})
should be substituted by the simplified ones:
\beq
\ov     {u}
        {\lambda + 2\eta \bar\epsilon_k}
        {\lambda}
        _ { j }
=\exp \left [-\frac{
2\pi j\i }{n} \big (u - n<\lambda,{\epsilon}_k>\big )
\right ] ,
\;\;\;\;\;\; j= 1, \ldots, n\, .
\label{expint}
\eeq
Eq.(\ref{vfc}) acquires the form
\begin{equation}
\sum_{i,j=1}^n
R(u-v)^{i,j}_{i',j'}\,\,
\ov     {u}
        {\mu}
        {\lambda}
        _{i}
\,\,
\ov     {v}
        {\nu}
        {\mu}
        _{j}
=
\sum_{\mu'}
\ov     {v}
        {\mu'}
        {\lambda}
        _{j'}
\,\,
\ov     {u}
        {\nu}
        {\mu'}
        _{i'}
\,\,
W
\left[\begin{array}{ccc}
                        & \mu & \\
                \lambda & u-v &\nu\\
                        & \mu'& \\\end{array}\right] .
\label{vfc1}
\end{equation}
The vertex model is now associated with the
simplified non-standard trigonometric $R$-matrix
$\Rmt$ (\ref{nrm1}) while the face model is the same as
in eq.\,(\ref{vfc}). The direct proof of formula (\ref{vfc1})
is given in Appendix B.

In the case $n=2$ both the 7-vertex and
6-vertex models are connected with the same face model
(\ref{facew1})-(\ref{facew3}) via the vertex-face
transformations (\ref{vfc}) and (\ref{vfc1}), respectively.
The \ivs are different in the two cases. To avoid a
confusion, we remark that in the paper \cite{deVega}
the 6-vertex $R$-matrix was related to a face model with
{\it constant} (i.e., independent of $\lambda$) face weights
by a transformation similar to (\ref{vfc1}). However, the
intertwining vectors in that transformation differ from
ours. In our scheme, that case corresponds to the {\it second}
scaling limit discussed in Sect.\,2 while our version
(\ref{vfc1}) corresponds to the {\it first} one.

The \ivs (\ref{expint})
can be simplified further. In fact their dependence on the
spectral parameter is irrelevant.
It can be eliminated by a gauge transformation.
This is achieved by the
transformation
\beq
R_{12}(u-v)\rightarrow  A_1(u) A_2(v) R_{12}(u-v)
( A_1(u) A_2(v))^{-1} :=R'_{12}(u-v)
\label{shift}
\eeq
with the diagonal matrix
$$A(u)_{ij}= \delta _{ij} \exp \left [
\frac{ \pi \i u }{n} (n-2j) \right ].$$
Under this transformation the intertwining vectors
loose their spectral parameters:
\beq
\ov     {u}
        {\lambda + 2\eta \bar\epsilon_k}
        {\lambda}
        _ { j }
\rightarrow
\tov     {\lambda + 2\eta \bar\epsilon_k}
        {\lambda}
        _ { j }
:=
\exp \big [ 2\pi \i j \, <\lambda,{\epsilon}_k> \big ] ,
\;\;\;\;\;\; j= 1, \ldots, n\, .
\label{tint}
\eeq
The $R$-matrix $R'(u)$ (\ref{shift}) has the form \cite{SU}, \cite{ABB}
\begin{eqnarray}
R'(u) ^{ij}_{i'j'}
&=&
\delta _{i,j} \delta_{i,i'} \delta _{i',j'}
\frac{\sin\pi(u+2\eta)}{ \sin 2\pi \eta }
\nonumber \\
&+&\delta_{i,i'} \delta _{j,j'} \,
\varepsilon (i'\neq j') \,
\frac{\sin \pi u}{ \sin 2\pi\eta}
\exp \left [ \frac{2\pi \i \eta}{n}
\left ( 2 (j-i ) - n \sign (j-i)\right )\right ]
\nonumber \\
&+&\delta _{i,j'} \delta _{i',j}
\, \varepsilon (i'\neq j')\,
\exp \big [\sign (j-i) \pi \i u \big ]
\nonumber \\
&-&2 \i \sin \pi u \,
\delta_{i+j,i'+j'}\,\varepsilon (i'<i<j')\,
\exp \left [\frac{
4\pi \i \eta }{n} (j'-i) \right ]
\nonumber \\
&+&2 \i \sin \pi u\,
\delta_{i+j,i'+j'}\,\varepsilon (i'>i>j')\,
\exp \left [\frac{
4\pi \i \eta }{n} (j'-i ) \right ].
\label{nrm2}
\end{eqnarray}

The \vfc for \ivs without spectral parameter reads
\begin{equation}
\sum_{i,j=1}^n
R'(u)^{i,j}_{i',j'}\,\,
\tov    {\mu}
        {\lambda}
        _{i}
\,\,
\tov    {\nu}
        {\mu}
        _{j}
=
\sum_{\mu'}
\tov     {\mu'}
        {\lambda}
        _{j'}
\,\,
\tov    {\nu}
        {\mu'}
        _{i'}
\,\,
W
\left[\begin{array}{ccc}
                        & \mu & \\
                \lambda & u &\nu\\
                        & \mu'& \\\end{array}\right].
\label{vfc2}
\end{equation}

\section{\,Factorized\, $L$-opera\-tor \, for\, the\, trigo\-nomet\-ric
Ruijse\-naars \,model}

Let us construct $L$-operators that satisfy the commutation
relation
\beq
R_{12}(u-v) L_1 (u) L_2 (v)
= L_2 (v) L_1 (u) R_{12}(u-v)
\label{yb}
\eeq
with the non-standard trigonometric $R$-matrices
(\ref{nrm}), (\ref{nrm1}) and (\ref{nrm2}).
As soon as the intertwining vectors are known,
this is straightforward. There is an important family
of $L$-operators made out of the intertwining vectors.
They are called
{\it factorized $L$-operators} \cite{H1}, \cite{QF}.
The name comes from a factorized form in which they are
presented.
It turns out that the factorized $L$-operators
serve as quantum Lax operators for the
trigonometric Ruijsenaars model.

Let us begin with the most general trigonometric $R$-matrix
(\ref{nrm}), (\ref{S}) and the corresponding intertwining
vectors (\ref{trint}). There are $n$ different vectors
labeled by $\bar \epsilon _{k}$, $k=1, \ldots , n$.
It is convenient to gather them in the $n\times n$ matrix
$\tilde \Phi ^{\lambda}(u)$
with matrix elements
\beq
(\tilde \Phi ^{\lambda}(u))_{jk}=
\tilde \ov
        {u}
        {\lambda+2\eta\bar{\epsilon_k}}
        {\lambda}
        _{j}.
\label{tPhi}
\eeq

The factorized $L$-operator reads
\beq
L^{(F)}(c|u) =
:\tilde \Phi ^{\lambda}(u+c)\cdot \hat T  \cdot
(\tilde \Phi ^{\lambda} (u))^{-1}:\, .
\label{flop}
\eeq
Here $\tilde \Phi ^{-1}$ denotes the inverse matrix,
the dot means matrix product, $c$ is a
parameter and $\hat T= \mbox{diag}\, ( T_1 , T_2 , \ldots , T_n )$
is
the diagonal operator matrix whose matrix elements are shift
operators:
$T_k f(\lambda)=  f(\lambda+2\eta\bar{\epsilon_k})$.
The normal ordering :: means that the shift operators
should be moved to the right after performing the matrix
product (cf. (\ref{t19})).

The $L$-operator (\ref{flop}) satisfies
eq.\,(\ref{yb}) with the
$R$-matrix (\ref{nrm}) for any parameter $c\in \C$. The proof
is based on the \vfc (\ref{vfc}) (cf. \cite{H2}).

It is useful to write down the factorized $L$-operator in
a slightly decoded form. For that purpose, introduce
components $\lambda _{k}=<\lambda , \epsilon _{k}>$
of the vector $\lambda$. These very parameters are to be
identified
with coordinates of particles in the Ruijsenaars model.
Matrix elements of $L^{(F)}(u)$
are difference operators acting to functions
$f(\lambda _{1}, \ldots , \lambda _{n})$:
\beq
L^{(F)}_{ij}(c|u)f(\lambda _{1}, \ldots , \lambda _{n})
=\sum _{k=1}^{n}
(\tilde \Phi ^{\lambda}(u+c))_{ik}
(\tilde \Phi ^{\lambda} (u))^{-1}_{kj}
T_k
f(\lambda _{1}, \ldots , \lambda _{n}),
\label{flop1}
\eeq
where
\begin{eqnarray*}
&&T_k
f(\lambda _{1}, \ldots , \lambda _{n})
\nonumber \\
&&=f( \lambda _{1}-\textstyle{ \frac{2\eta}{n}},\,  \ldots , \,
\lambda _{k-1}-\textstyle{\frac{2\eta}{n}},\,
\lambda _{k}+\textstyle{\frac{2\eta (n-1)}{n}},\,
\lambda _{k+1}-\textstyle{\frac{2\eta}{n}},\,
\ldots ,\,
\lambda _{n}-\textstyle{\frac{2\eta}{n}}).
\end{eqnarray*}
Note that $T_k$ commutes with the sum of coordinates
$\sum \lambda _{k}$.

In a similar way, we find the factorized $L$-operator
for the $R$-matrix (\ref{nrm1}). It is given by
the same formula (\ref{flop}) with the matrix
\beq
(\Phi ^{\lambda}(u))_{jk}=
\ov
        {u}
        {\lambda+2\eta\bar{\epsilon_k}}
        {\lambda}
        _{j}
\label{Phi}
\eeq
in place of $\tilde \Phi$. Here, the intertwining vectors
are as in (\ref{expint}).

Finally, for the spectral parameter independent version
(\ref{tint}) we introduce the matrix
\beq
\Phi ^{\lambda}_{jk}=
\tov     {\lambda + 2\eta \bar\epsilon_k}
        {\lambda}
        _ { j }
=
\exp \big [ 2\pi \i j \, <\lambda,{\epsilon}_k> \big ].
\label{expPhi}
\eeq
The factorized $L$-operator takes the simple form
\beq
L =
:\Phi ^{\lambda -\kappa}\cdot \hat T \cdot
(\Phi ^{\lambda})^{-1}:\, ,
\label{flop2}
\eeq
where
$\kappa \equiv \frac{c}{n}\sum _{k=1}^{n}\epsilon _{k}$.
It does not depend on the spectral parameter and satisfies
\beq
R_{12}' (u-v) L_1 L_2 = L_2 L_1 R_{12}' (u-v)
\label{yb2}
\eeq
with the $R$-matrix (\ref{nrm2}). These formulas generalize
eqs\,(\ref{t23})-(\ref{t25}) to the case $n\geq 3$.

Tending $u - v$ to $- \i \infty$, it is possible to get rid of the
spectral parameter in the $R$-matrix, too:
\begin{eqnarray}
{R_{CG}}^{ij}_{i'j'}
&=&
\delta _{i,j} \delta_{i,i'} \delta _{i',j'}
\frac{ \exp \big ( 2 \pi \i \eta \big )}{ \sin 2\pi \eta }
\nonumber \\
&+&\delta_{i,i'} \delta _{j,j'}\,
\varepsilon (i'\neq j')\,
\frac{1}{ \sin 2\pi\eta}
\exp \left [ \frac{
2\pi \i \eta}{n}
\big ( 2 (j-i )- n\sign(j-i) \big )\right ]
\nonumber \\
&+&\delta _{i,j'} \delta _{i',j}
\, 2\i \, \varepsilon (i'> j')
\nonumber \\
&-&2\i \,
\delta_{i+j,i'+j'}\,\varepsilon (i'<i<j' )\,
\exp \left [
\frac{ 4\pi\i \eta }{n} \big ( j'-i \big )\right ]
\nonumber \\
&+&2\i \,
\delta_{i+j,i'+j'}\,\varepsilon (i'>i>j' )\,
\exp \left [
\frac{ 4\pi\i \eta }{n} \big ( j'-i \big )\right ].
\label{constnrm}
\end{eqnarray}
This $R$-matrix coincides with the one given in (\ref{constnrm}).
It satisfies the Yang-Baxter equation
without spectral parameter \cite{FRT}. For $n=2$, eq.\,(\ref{yb2})
encodes commutation relations of the algebra of functions on the
quantum group $SL(2)$.

Taking trace of any one of the constructed $L$-operators,
we get, up to an irrelevant common factor,
\beq
M_{1}
=
\sum_{k=1}^{n}
\left (
\prod _{j=1, j\neq k}^{n}
\frac
    {\sin \pi( \lambda _{jk} + \frac{c}{n})}
    {\sin \pi \lambda _{jk}}
\right )
  T_k.
\label{im1}
\eeq
This operator is gauge equivalent to the first non-trivial
Hamiltonian of the trigonometric $n$-particle Ruijsenaars model.

Applying the fusion procedure to the factorized $L$-operators,
it is possible to obtain the whole commutative family
of IM as traces of the fused
$L$-operators (see \cite{H2} for details). In this way
one gets the commuting Macdonald operators \cite{McD}
\beq
M_{d}
=
\sum_{I\subset \{1, \ldots , n\}, |I|=d}
\left (
\prod _{r\notin I, s\in I}
\frac
    {\sin \pi( \lambda _{rs} + \frac{c}{n})}
    {\sin \pi \lambda _{rs}}
\right )
  T_I, \;\;\;\;\;\;d=1, \ldots , n-1,
\label{imd}
\eeq
where $T_I=\prod_{i\in I} T_i$. By a conjugation with a
function of $\lambda$ they yield higher commuting Hamiltonians
of the trigonometric Ruijsenaars model.

The generating function for these IM is
\beq
:\det [L - z]:=
\sum_{d=0}^n (-z)^{n-d} M_d ,
\label{sim}
\eeq
where $M_0 =1$, $M_n = \prod _{k=1}^{n} T_{k}$.
In the elliptic case, a similar formula was proved in \cite{H2}.

\section{Connection with the dynamical $R$-matrix approach}

A more familiar approach to the Ruijsenaars-like models is
based on dynamical $R$-matrices.
Let us show how to get the dynamical $R$-matrix  \cite{F}, \cite{ABB}
\begin{eqnarray}
R^{D}(u,\lambda )_{r\,s}^{r'\,s'}
&=&
\delta_{r r'}\delta_{s s'}\delta_{r s}
\frac {\sin \pi(u+{2\eta})}{\sin {2\pi \eta} }
\nonumber \\
&+&
\delta_{r s'}\delta_{r' s} \, \varepsilon(r\neq s) \,
\frac{\sin \pi (-u+\lambda_{rs})}{\sin \pi \lambda_{rs}}
\nonumber \\
&+&
\delta_{r r'}\delta_{s s'}\, \varepsilon(r\neq s)\,
\frac{\sin \pi u}{\sin {2 \pi \eta}}
        \frac{\sin \pi({2\eta}+\lambda_{rs})}
{\sin \pi \lambda_{rs}}
\label{qdrm}
\end{eqnarray}
from the non-standard \trig one (\ref{nrm1})
by a quasi-Hopf twist \cite{D}.

Consider the \vfc (\ref{vfc1}) with the face weights
(\ref{facew1})-(\ref{facew3}) and the $R$-matrix (\ref{nrm1}).
By $\Phi_1^{\lambda}(u)$ denote
the matrix  of intertwining vectors
(\ref{Phi}) in the first copy of the space ${\bf C}^n$:
$$
(\Phi ^{\lambda }(u))_{i'}^{i}=
\ov     {u}
        {\lambda + 2\eta \bar \epsilon_i}
        {\lambda}
        _ { i'}
$$
and by $\Psi _{12}^{\lambda}(u)$ denote the following matrix
in the tensor product
${\bf C}^n \times {\bf C}^n$:
$$
(\Psi^{\lambda}(u))_{i'j'}^{ij}=\delta _{ii'}
\ov     {u}
        {\lambda  + 2\eta \bar \epsilon_i+ 2\eta \bar \epsilon_j}
        {\lambda  + 2\eta \bar \epsilon_{i'}}
        _ { j'}
$$
(it is a diagonal matrix in the first copy of
${\bf C}^n$).

In this notation, the vertex-face correspondence (\ref{vfc1})
can be written as a matrix equation:
\beq
R_{12}(u-v)
\Phi_1^{\lambda}(u)
\Psi _{12}^{\lambda}(v)=
\Phi_2^{\lambda}(v)
\Psi _{21}^{\lambda}(u)
W_{12}(u-v,\lambda )
\label{svfc}
\eeq
For convenience, we write down the same equation with indices:
\beq
\sum _{i,j,k}
R(u-v)_{i'j'}^{ij}
(\Phi^{\lambda}(u))_{i}^{k}
(\Psi ^{\lambda}(v))_{kj}^{mr}=
\sum _{i,j,r'}
(\Phi^{\lambda}(v))_{j'}^{r'}
(\Psi ^{\lambda }(u))_{i'r'}^{ij}
W(u-v,\lambda )_{ij}^{mr}\,.
\label{svfc1}
\eeq
Here the matrix $W_{12}(u,\lambda)$ is
$$
(W (u,\lambda ))_{r'\,s'}^{r\,s}
=\delta_{r+s,r'+s'}\,\,
W
\left[\begin{array}{ccc}
                & \lambda +2\eta\bar{\epsilon_{r}} & \\
        \lambda &u&\lambda + 2\eta(\bar{\epsilon_{r}}
        +\bar{\epsilon_{s}})\\
                & \lambda +2\eta\bar{\epsilon_{s'}} &
                \\\end{array}\right]
$$
which coincides with the dynamical $R$-matrix (\ref{qdrm}):
\beq
W_{12}(u, \lambda ) = R_{12}^{D}(u, \lambda) \,.
\label{WR}
\eeq
The $W$-weights are matrix elements
of the dynamical $R$-matrix.

We conclude that the dynamical and non-dynamical
$R$-matrices are connected by a quasi-Hopf twist:
\beq
R_{12}(u-v) F_{12}(u,v;\lambda)=
F_{21}(v,u;\lambda ) R_{12}^{D}(u-v, \lambda)\,,
\label{twist}
\eeq
where
\beq
F_{12}(u,v;\lambda )=\Phi _{1}^{\lambda }(u)
\Psi _{12}^{\lambda }(v)
\,.
\label{F12}
\eeq
Thus we have represented the \vfc (\ref{vfc1}) in the form
of a quasi-Hopf twist connecting the non-dynamical
$R$-matrix with a dynamical one\footnote{The quasi-Hopf
relations (\ref{twist}), (\ref{F12})
are also valid for the \vfc (\ref{vfc}) with non-standard \trig $R$-matrix
(\ref{nrm}), (\ref{S}) and the \ivs (\ref{trint}).}.
The matrix $F_{12}$ in the form (\ref{F12}) (up to a factor
commutative with the $R$-matrix) was calculated for $n=2$
in \cite{Bab}.

It is known \cite{ABB}
that the dynamical $R$-matrix (\ref{qdrm})
is also related to the standard \trig $R$-matrix (\ref{srm})
by another quasi-Hopf twist:
\beq
R_{A_{n-1}}(u)_{12} \tilde F_{12}(\lambda )
=\tilde F_{21}(\lambda ) R_{12}^{D}(u,\lambda )\,,
\label{htwist}
\eeq
\begin{eqnarray*}
\tilde F(\lambda )_{r\,s}^{r's'}
&=& 2\i  \,
\delta_{r r'}\delta_{s s'}\delta_{r s}
\nonumber \\
&+&
\delta_{r r'}\delta_{s s'}\, \varepsilon (r < s)\,\,
\frac{1}
{\sin \pi \lambda_{rs}},
\nonumber \\
&+&
\delta_{r r'}\delta_{s s'}\, \varepsilon (r >s)\,\,
\frac{1}
{\sin \pi (2\eta-\lambda_{rs})},
\nonumber \\
&-&
\delta_{r s'}\delta_{r' s} \, \varepsilon (r < s)\,\,
\frac{\exp \left[\pi\i\lambda_{rs}\right] \sin 2\pi\eta }
{\sin \pi \lambda_{rs} \sin \pi (2\eta+\lambda_{rs})}
\nonumber \\
\end{eqnarray*}

Comparing the relations (\ref{twist}) and  (\ref{htwist}),
one has
\beq
R_{12}(u-v)={F'}_{21}(v,u ;\lambda ) R_{A_{n-1}}(u-v)_{12}
(F'_{12}(u,v; \lambda ))^{-1}
\label{qhopf}
\eeq
with
$$
F'_{12}(u,v;\lambda )=F_{12}(u,v; \lambda )
(\tilde F_{12}(\lambda ))^{-1}\,.
$$
Therefore, the two {\it non-dynamical} $R$-matrices turn out
to be related by a quasi-Hopf twist which depends on the dynamical
variables.

\section{Conclusion}

Let us summarize the results. Starting from the elliptic
$n^2$$\times$$n^2$ Belavin $R$-matrix
$R^{(ell)}(u)$,
we have considered the chain
$$
\tilde R (u)\,\, \rightarrow \,\,
R (u)\,\, \rightarrow \,\,
R'(u)\,\, \rightarrow \,\,
R_{CG}
$$
of {\it non-standard} trigonometric $R$-matrices obtained
as its different degenerations
and given by (\ref{nrm}), (\ref{nrm1}), (\ref{nrm2}) and
(\ref{constnrm}), respectively. We call them non-standard
because they differ from the standard trigonometric $R$-matrix
$R _{A_{n-1}}(u)$ (\ref{srm}).
The arrows mean certain types
of degeneration procedures described in the main body of the
paper. The last $R$-matrix in this chain does not depend on
the spectral parameter and coincides with the constant
$R$-matrix introduced by Cremmer and Gervais \cite{CG} in
another context.

There are two things common for all these
$R$-matrices:
\begin{itemize}
\item  They satisfy the standard Yang-Baxter equation
\item  They are {\it non-dynamical} $R$-matrices for the
$n$-particle trigonometric Ruijsenaars model.
\end{itemize}

The quantum Lax matrices for the trigonometric Ruijsenaars
model intertwined by these $R$-matrices have been constructed
using the technique of intertwining vectors. In each case,
the Hamiltonian of the model is obtained as trace of the Lax
matrix up to a common non-essential factor.

It is surprising that the standard trigonometric $R$-matrix
$R _{A_{n-1}}(u)$ {\it does not} belong to this chain (except
for the case $n=2$) in the sense that it is not appropriate for
the Ruijsenaars-type models. Though, it is obtained as a
limiting case of the Belavin $R$-matrix, too, in a way around
the chain. It has been also shown that the non-standard
$R$-matrices (\ref{nrm}), (\ref{nrm1}) can be obtained
from the standard one (\ref{srm}) by a quasi-Hopf twist.

A more customary (and until the very
recent time the only available) $R$-matrix formulation of
the Ruijsenaars-type models is based on {\it dynamical} $R$-matrices.
The explicit connection between the dynamical and non-dynamical
$R$-matrices is given by eq.\,(\ref{twist}) that has the form
of Drinfeld's quasi-Hopf twist. At the same time this
is a simple reformulation of the famous vertex-face correspondence
between "vertex" and "face" type lattice statistical models
with trigonometric Boltzmann weights. The dynamical $R$-matrix
is identified with the matrix of $W$-weights for the face model.

\section*{Acknowledgments}

We are grateful to A.Alekseev, G.Arutyunov, J.Avan, A.Belavin,
E.Billey, L.Chekhov, H.J. de Vega, B.Enriquez, L.Faddeev,
B.Feigin, S.Frolov, P.Kulish, J.M.Maillard,
T.Miwa, N.Reshetikhin, S.Sergeev,
F.Smirnov, Yu. Stroganov, C.M. Viallet
for fruitful discussions and comments.
A.A. is thankful to L.Baulieu and
Laboratoire de Physique Th\'eorique et Hautes \'Energies,
Paris, France for the hospitality.
A.A. also would like to thank O.Babelon
for the great help and interest to this work.
A.A. thanks M.Coulon for encouragement.
Some results of this work were reported
by the first author in the talk given
on the Semester "Syst\`emes int\'egrables",
15 septembre 96 - 14 f\'evrier 97,
Centre Emile Borel, Institut Henri Poincare, Paris, France.
The work of A.Z. was supported in part by
RFBR grant 97-02-19085 and by ISTC grant 015.
K.H. acknowledges the hospitality 
at Centre de Recherche 
Mathematique and Universit{\'e} de Montr{\'e}al 
during his 
visit and especially thanks Professor J.F. van Diejen.

\section*{Appendix A}

In this Appendix we show how to get the twisted trigonometric
limit of the intertwining vectors. We use the theta
functions with rational characteristics (\ref{thetaj}).
We also need the Jacobi theta function
$$
\theta _{1}(u)=  \sum_{ m\in {\bf Z}}
        \exp \left (\pi i \tau (m+\half)^2  +
2\pi i (m+\half)(u+\half )\right ).
$$
The elliptic face weights are \cite{JMO}, \cite{H1}:
$$
W^{(ell)}
\left[\begin{array}{ccc}
                        & \lambda +2\eta\bar{\epsilon_{i}} & \\
                \lambda &u& \lambda + 4\eta\bar{\epsilon_{i}} \\
                        & \lambda +2\eta\bar{\epsilon_{i}} &
                        \\\end{array}\right
]
= \frac {\theta_1(u+{2\eta})}{\theta_1({2\eta})},
$$
$$
W^{(ell)}
\left[\begin{array}{ccc}
                & \lambda +2\eta\bar{\epsilon_{i}} & \\
        \lambda
        &u&\lambda+2\eta(\bar{\epsilon_{i}}+\bar{\epsilon_{j}})\\
                & \lambda +2\eta\bar{\epsilon_{i}} &
                \\\end{array}\right]
= \frac {\theta_1(-u+\lambda_{ij})}{\theta_1(\lambda_{ij})},
\;\;\;\; i\neq j,
$$
$$
W^{(ell)}
\left[\begin{array}{ccc}
                & \lambda +2\eta\bar{\epsilon_{i}} & \\
        \lambda &u&\lambda + 2\eta(\bar{\epsilon_{i}}
        +\bar{\epsilon_{j}})\\
                & \lambda +2\eta\bar{\epsilon_{j}} &
                \\\end{array}\right]
= \frac {\theta_1(u)}{\theta_1({2\eta})}
        \frac {\theta_1({2\eta}+\lambda_{ij})}
{\theta_1(\lambda_{ij})},
\;\;\;\; i\neq j.
$$
The vertex-face correspondence
is implemented by elliptic intertwining vectors
$$
\ovell{u}
        {\mu}
        {\lambda}
        _j
=
\left\{\begin{array}{cl}
        \theta ^{j} \big (u-n<\lambda , \bar \epsilon _{k} >
+\textstyle{ \frac{n-1}{2} }\big )

        & :\mu-\lambda=2\eta\bar{\epsilon}_k
                \quad \hbox{ for some}\;\; k=1,\ldots ,n
,\\
        0
        & : \hbox{ otherwise}\\
\end{array}\right.
$$
The elliptic vertex-face correspondence has the form
\begin{eqnarray*}
&&\sum_{i,j=1}^n
R^{(ell)}(u-v)^{i,j}_{i',j'}
\ovell   {u}
        {\mu}
        {\lambda}
        _{i}
\ovell     {v}
        {\nu}
        {\mu}
        _{j}
\nonumber \\
&=&
\rho (u) \sum_{\mu'}
\ovell     {v}
        {\mu'}
        {\lambda}
        _{j'}
\ovell     {u}
        {\nu}
        {\mu'}
        _{i'}
W^{(ell)}
\left[\begin{array}{ccc}
                        & \mu & \\
                \lambda & u-v &\nu\\
                        & \mu'& \\\end{array}\right],
\end{eqnarray*}
where $\rho (u)$ is a normalization factor such that
$\rho (u) \rightarrow 1$ as $h \rightarrow 0$, $h=e^{\pi \i \tau}$.

Taking the limit
or $h\rightarrow 0$ of this equation and using formulas
given in Sect.\,3,
we see that \ivs diverge. To regularize them,
let us extract the singular $h$-dependent factors:
$$ \ovell    {u}
        {\mu + 2\eta \bar\epsilon_k}
        {\mu}
        _ { j }
=
h^{n(\half-\frac{j}{n})^2}
\Ovell     {u}
        {\mu + 2\eta \bar\epsilon_k}
        {\mu}
        _ { j }
\;\; \hbox { for }\;\; j= 1, \ldots, n\, ,
$$
where
$
\Ovell    {u}
        {\mu + 2\eta \bar\epsilon_k}
        {\mu}
        _ { j }
$
already have a smooth limit $h \rightarrow 0$ coinciding with
vectors
$\tilde \ov     {u}
        {\mu + 2\eta \bar\epsilon_k}
        {\mu}
        _ { j }
+\sum_{\alpha>0} h^\alpha \phi_\alpha
$
introduced in eq.\,(\ref{trint}).
Rewriting the elliptic vertex-face correspondence
as
\begin{eqnarray*}
&{}&\sum_{i,j=1}^n
\left( R^{(ell)}(u-v)^{i,j}_{i',j'}
h^{-n(\half-\frac{j'}{n})^2
-n(\half-\frac{i'}{n})^2
+n(\half-\frac{j}{n})^2
+n(\half-\frac{i}{n})^2
}\right)
\Ovell     {u}
        {\mu}
        {\lambda}
        _{i}
\Ovell    {v}
        {\nu}
        {\mu}
        _{j}
\\
&{}&= \rho (u)
\sum_{\mu'}
\Ovell     {v}
        {\mu'}
        {\lambda}
        _{j'}
\Ovell    {u}
        {\nu}
        {\mu'}
        _{i'}
W^{(ell)}
\left[\begin{array}{ccc}
                        & \mu & \\
                \lambda & u-v &\nu\\
                        & \mu'& \\\end{array}\right]
\end{eqnarray*}
and taking the limit, we come to the trigonometric
version (\ref{vfc}.

\section*{Appendix B}

Here we give a direct proof of eq.\,(\ref{vfc2}).

Specify the face variables as
$$
\mu=\lambda + 2\eta\bar\epsilon_r \hbox{ and }
\nu=\mu+ 2\eta\bar\epsilon_s
\,\,\,\,\, \hbox { for some }\;\; r,s=1, \ldots n.
$$

One starts with the case
$i'=j'$.
The $R$-matrix in the l.h.s. of (\ref{vfc2}) is equal to
$$
\delta _{i,j} \delta_{i,i'} \delta _{i',j'}
\frac{\sin\pi(u+2\eta)}{ \sin 2\pi \eta }.
$$
Let $r=s$. Then we have the only term in the r.h.s.
with the $W$-weight (\ref{facew1}). The \vfc (\ref{vfc2})
is evident in this case.

Now let  $r\neq s$. The l.h.s. of (\ref{vfc2}) reads
$$
\frac{\sin\pi(u+2\eta)}{ \sin 2\pi \eta }
\exp \left [2\pi \i j'
\big (<\lambda, \epsilon_s+ \epsilon_r> -\frac{2\eta}{n}\big )
\right ].
$$
In the r.h.s. one has two terms corresponding to
$W$-weights (\ref{facew2}) and (\ref{facew3}):
\begin{eqnarray*}
&{}&
\frac {\sin\pi(-u+\lambda_{rs})}{\sin \pi \lambda_{rs}}
\exp \left [
2\pi \i j'(<\lambda, \epsilon_s+ \epsilon_r> -\frac{2\eta}{n})
\right ]
\\
&+&\frac {\sin\pi u}{\sin 2\pi \eta}
        \frac {\sin\pi({2\eta}+\lambda_{rs})}{\sin\pi \lambda_{rs}}
\exp \left [
2\pi \i j'(<\lambda, \epsilon_s+ \epsilon_r> -\frac{2\eta}{n})
\right ].
\end{eqnarray*}
Eq.\,(\ref{vfc2}) follows from the trigonometric
identity
$$
\frac{\sin\pi(u+2\eta)}{ \sin 2\pi \eta }
=\frac {\sin\pi(-u+\lambda_{rs})}{\sin\pi \lambda_{rs}}+
\frac {\sin\pi({2\eta}+\lambda_{rs})}{\sin\pi \lambda_{rs}}.
$$

Let $i'<j'$.
Then the $R$-matrix in the l.h.s. of (\ref{vfc2}) takes the form
\begin{eqnarray*}
&{}&
\delta_{i,i'} \delta _{j,j'}
\frac{\sin \pi u}{ \sin 2\pi\eta}
\exp 2\pi \i \eta
\left[ \frac{2}{n} (j'-i') -1 \right]
\\
&+&\delta _{i,j'} \delta _{i',j}
\exp \left [-\pi \i u\right ]
\\
&-&2 \sin \pi u \,
\delta_{i+j,i'+j'}\,\varepsilon (i'<i<j')\,
\exp 2\pi\i\left[\frac{1}{4}+ \frac{2\eta }{n}
(j'-i )  \right].
\end{eqnarray*}
The l.h.s. of (\ref{vfc2}) reads
\begin{eqnarray*}
&{}&\frac{\sin \pi u}{ \sin 2\pi\eta}
\exp 2\pi \i \eta
\left[ \frac{2}{n} (j'-i') -1 \right]
\,\,
\tov    {\mu}
        {\lambda}
        _{i'}
\,\,
\tov    {\nu}
        {\mu}
        _{j'}
+
\exp \big [-\pi \i u \big ]
\tov    {\mu}
        {\lambda}
        _{j'}
\,\,
\tov    {\nu}
        {\mu}
        _{i'}
\\
&-&2 \sin \pi u \sum_{1\leq i' < i <j'\leq n}
\,
\exp 2\pi\i\left[\frac{1}{4}+\frac{2\eta }{n} (j'-i )  \right]
\,\,
\tov    {\mu}
        {\lambda}
        _{i}
\,\,
\tov    {\nu}
        {\mu}
        _{i'+j'-i} .
\end{eqnarray*}
After some algebra the l.h.s.
becomes
\begin{eqnarray*}
\hbox{ l.h.s } &=&
\frac{\sin \pi u}{ \sin 2\pi\eta}
\exp 2\pi \i
\left[ i'(<\lambda, \epsilon_r>-\frac{2\eta}{n})
+j' (<\lambda, \epsilon_s>+ 2\eta\delta_{rs})-\eta \right]
\\
&+& \exp \pi \i (-u) \,\,\exp 2\pi \i
\left[ j'  <\lambda, \epsilon_r> +
i' (<\lambda, \epsilon_s> +2\eta (\delta_{rs}-\frac{1}{n}))\right]
\\
&-&2 \sin \pi u \exp 2\pi\i \left[\frac{1}{4}
+ i'(<\lambda,\epsilon_s>+ 2\eta(\delta_{rs}-\frac{1}{n}))
+  j'(<\lambda,\epsilon_s>+ 2\eta\delta_{rs} )   \right]
\\
&\times&
\frac{ \exp 2\pi \i \left[(\lambda_{rs}- 2\eta\delta_{rs})(i'+1) \right]
-\exp 2\pi \i \left[(\lambda_{rs}- 2\eta\delta_{rs}) \,\,j' \right]
}
{1- \exp 2\pi \i(\lambda_{rs}- 2\eta\delta_{rs})}.
\end{eqnarray*}
Now specify the indexes $r,s$. First, let
$r=s$, then the l.h.s. is
\begin{eqnarray*}
&{}&\hbox{ l.h.s } =
\exp 2\pi\i\left[
i'(<\lambda,\epsilon_r>-\frac{2\eta}{n})
+  j' \,<\lambda,\epsilon_r> \right]
\cr
&\times&
\left\{
\frac{\sin \pi u}{ \sin 2\pi\eta}
\exp 2\pi \i \left[ 2 j' \eta-\eta     \right]
+
\exp \pi \i (-u) \exp [ 4\pi \eta \i i' \, ]
\right.
\cr
&-& \left.
\frac{\sin \pi u}{\sin 2\pi\eta}
 \exp 2\pi\i
\left[
2i' \eta+ 2j' \eta
\right]
\left( \exp 2\pi \i \left[- 2 i'\eta -\eta\right]
-\exp 2\pi \i \left[- 2 j'\eta  +\eta \right]
\right)
\right\}
\cr
\end{eqnarray*}
In the case $r=s$ there is only one term in the r.h.s.
of (\ref{vfc2}) corresponding to the face weight (\ref{facew1}):
$$
\hbox{ r.h.s } =
\exp 2\pi\i\left[
 i'\,(<\lambda,\epsilon_r>+2\eta (1-\frac{1}{n}))
+  j' \,<\lambda,\epsilon_r> \right]
\frac {\sin\pi(u+{2\eta})}{\sin\pi({2\eta})} .
$$
The equality of the l.h.s. and
the r.h.s. follows from the identity
$$
\exp [-\pi \i u ]+\frac{\sin \pi u}{\sin 2\pi\eta}
\exp [2\pi\i \eta ]
=\frac {\sin\pi(u+{2\eta})}{\sin 2 \pi \eta}.
$$

Consider now the case $r\neq s$, then the l.h.s.
of eq.\,(\ref{vfc2}) after some simplifications reads
\begin{eqnarray*}
\hbox{ l.h.s } &=&
\frac{\sin \pi u}{ \sin 2\pi\eta}
\exp 2\pi \i
\left[ i'(<\lambda, \epsilon_r>-\frac{2\eta}{n})
+j' \,<\lambda, \epsilon_s>-\eta \right]
\\
&+& \exp \pi \i (-u) \,\,\exp 2\pi \i
\left[ j'  <\lambda, \epsilon_r> +
i' (<\lambda, \epsilon_s> -\frac{2\eta}{n}) \right]
\\
&-& \frac{\sin \pi u}{\sin\pi\lambda_{rs}}
\exp 2\pi\i \left[
 (i'+\half)<\lambda,\epsilon_s> - \frac{2i' \eta}{n}
+  (j'-\half)<\lambda,\epsilon_r>  \right].
\\
&+&
\frac{\sin \pi u}{\sin\pi\lambda_{rs}}
\exp 2\pi\i \left[
 (i'+\half)<\lambda,\epsilon_r> - \frac{2i' \eta}{n}
+  (j'-\half)<\lambda,\epsilon_s>  \right]
\end{eqnarray*}
In the r.h.s. there are two terms corresponding to
the $W$-weights (\ref{facew2}) and (\ref{facew3}):
\begin{eqnarray*}
\hbox{ r.h.s } &=&
\exp 2\pi\i \left[
 i'\,<\lambda,\epsilon_s> - \frac{2i' \eta}{n}
+ j'\,<\lambda,\epsilon_r>  \right]
\frac {\sin\pi(-u+\lambda_{rs})}{\sin \pi \lambda_{rs}}
\\
&+&
\exp 2\pi\i \left[
 i'\,<\lambda,\epsilon_r> - \frac{2i' \eta}{n}
+ j'\,<\lambda,\epsilon_s>  \right]
\frac {\sin \pi u}{\sin 2\pi \eta }
        \frac {\sin\pi({2\eta}+
\lambda_{rs})}{\sin \pi \lambda_{rs}}.
\end{eqnarray*}
It is easy to see that the \vfc  in this case
is equivalent to
$$
A\,\,\exp 2\pi\i \left[
 i'\,<\lambda,\epsilon_s>
+ j'\,<\lambda,\epsilon_r>  \right]
+ B\,\, \exp 2\pi\i \left[
 i'\,<\lambda,\epsilon_r>
+ j'\,<\lambda,\epsilon_s>  \right]
\equiv 0 ,
$$
where $A$ and $B$ are some trigonometric
functions of $\eta$, $u$ and $\lambda_{rs}$. Thus, to prove
the \vfc it is enough to show that $A\equiv 0$ and $B\equiv 0$.
The result follows from
$$
\frac{\exp [-2\pi\i \eta ]}{\sin 2\pi\eta}
+\frac{\exp [\pi\i \lambda_{rs}] }{\sin \pi \lambda_{rs}}
-\frac{\sin\pi(2\eta+\lambda_{rs})}
{\sin 2\pi\eta\,\, \sin\pi\lambda_{rs}}
\equiv 0
$$
and
$$
-\exp [-\pi\i u ]
+\frac{\sin\pi u}{\sin \pi\lambda_{rs}}
\exp [-\pi \i \lambda_{rs} ]+
\frac{\sin\pi(-u+\lambda_{rs})}{\sin\pi\lambda_{rs}}
\equiv 0.
$$
The case $i'>j'$ is considered similarly.

\end{document}